\documentclass[aps,pre,superscriptaddress,floatfix,showpacs]{revtex4}
\usepackage{graphicx}
\usepackage{amssymb}
\usepackage[usenames]{color}

\begin{document}

\title{How simple regulations can greatly reduce inequality}

\author{J. R. Iglesias}
\affiliation{Programa de P\'os-Gradua\c{c}\~ao em Economia and Instituto de F\'{\i}sica, Universidade Federal do Rio Grande do Sul, 91501-970 Porto Alegre, Brazil}

\date{\today}

\begin{abstract}

Many models of market dynamics make use of the idea of wealth exchanges among economic agents. A simple analogy compares the wealth in a society with the energy in a physical system, and the trade between agents to the energy exchange between molecules during collisions. However, while in physical systems the equipartition of energy is valid, in most exchange models for economic markets the system converges to a very unequal ``condensed'' state, where one or a few agents concentrate all the wealth of the society and the wide majority of agents shares zero or a very tiny fraction of the wealth. Here we present an exchange model where the goal is not only to avoid condensation but also to reduce the inequality; to carry out this objective the choice of interacting agents is not at random, but follows an extremal dynamics regulated by the wealth of the agent. The wealth of the agent with the minimum capital is changed at random and the difference between the ancient and the new wealth of this poorest agent is taken from other agents, so establishing a regulatory tool for wealth redistribution. We compare different redistribution processes and conclude that a drastic reduction of the inequality can be obtained with very simple regulations.
\end{abstract}


\maketitle

\section{Introduction}

Empirical studies focusing the income distribution of workers, companies and countries were first presented more than a century ago by Italian economist Vilfredo Pareto. He observed, in his book ``Cours d'Economie Politique''~\cite{Pareto}, that the distribution of income does not follow a Gaussian distribution but a power law. That means that the asymptotic behavior of the distribution function is not exponential (as it should be for a Gaussian distribution) but follows a power function that decreases, for big values of the wealth, as $w^{-\alpha}$, being $\alpha > 1$ the exponent of the power law. Non-gaussian distributions are denominated Levy distributions~\cite{Mantegna}, thus this power law distribution is nowadays known as Pareto-Levy Distribution. Then, Pareto asserted that in different countries and times the income distribution follows a power law behavior, i.e.  the cumulative probability $P(w)$ that an agents have an income which is at higher or equal to $w$ is given by $P(w) \propto w^{-\alpha}$~\cite{Pareto}. The exponent $\alpha$ is named Pareto index. The value of this exponent changes with geography and time, but a typical values are close to $3/2$. The bigger the value of the Pareto exponent the higher the inequality in a society.

However, recent data indicate that, even though Pareto distribution provides a good fit in the the high income range, it does not agree with the observed data over the middle and low income ranges. For instance, data from Japan~\cite{souma,nirei}, Italy~\cite{clementi}, India~\cite{sinha1}, the United States of America and the United Kingdom~\cite{dragu2000,dragu2001a,dragu2001b} are fitted by a log-normal or Gaussian distribution with the maximum located at the middle income region plus
a power law for the high income strata. This kind of behavior is also observed in the wage distribution in Brazil. We have represented on Fig. 1 the number of families as a function of the annual income and it is clear that for low and middle salaries the distribution follows an almost Gaussian law, while for high salaries the curve clearly departs from the exponential and may be described, in spite of the high noise, as a power law distribution. The Pareto exponent for this distribution is $2.7$. If we consider that the results in Fig. 1 are not cumulative, The integration of the curve will deliver the cumulative distribution and the high income region will show a exponent $1.7$ very near Pareto's value of $3/2$.

One way of justifying in a qualitative way the two regimes: Gaussian $+$ power law, is by considering that in the low and middle income classes the process of wealth accumulation is additive, causing a Gaussian-like distribution, while in the high income range, wealth grows in a multiplicative way (interests or other mechanism of multiplying assets), generating the observed power law tail~\cite{nirei}.

Another verification of Pareto-Levy's law is shown on Fig. 2, where the accumulated Gross Domestic Product of the $100$ richest cities in Brazil~\cite{IBGE2} is plotted as a function of the ranking. The county of S\~ao Paulo concentrates the biggest fraction of the total Brazilian domestic product, around $12\%$, followed by Rio de Janeiro, Bras\'{\i}lia and Belo Horizonte.
All in all the first $100$ counties in the ranking concentrate almost $60 \%$ of the Brazilian GDP. The log-log plot used on Fig. 2 makes the power law explicit , as a power function becomes a straight line when plotted in double logarithmic scale (while exponentials, or Gaussians, look as parabolas, for example the low -- middle income region in Fig. 1). This power law is valid for the richest counties. However, as there are more than $5500$ in Brazil, the vast majority (not included in the graphic) shares the other $40 \%$ of the GDP, so, it is certain that the distribution deviates form the power low and becomes a Gaussian distribution for the big majority of cities and towns.
\begin{figure}
\includegraphics[width=0.7\textwidth]{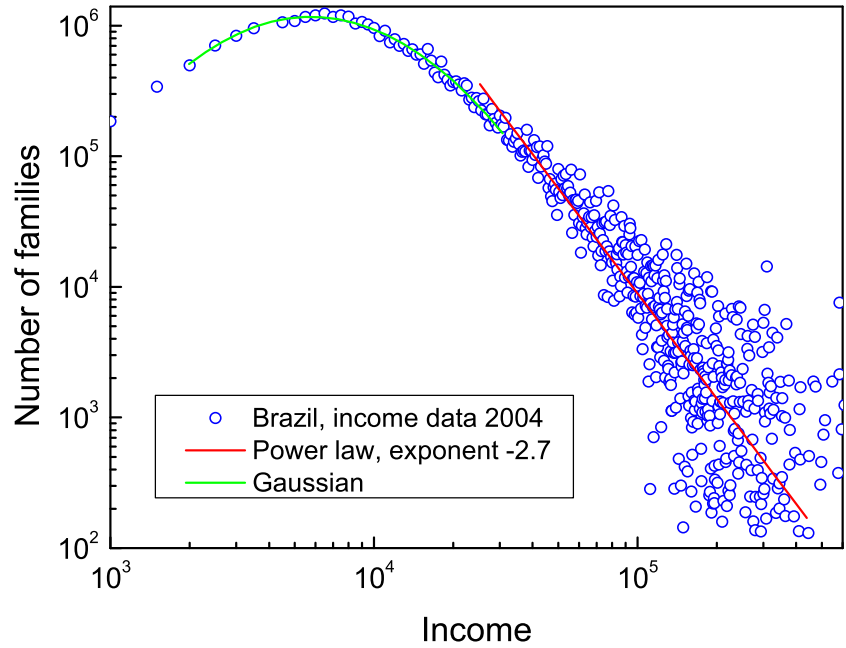}
\caption{Number of families per salary class (in Brazilian Real, annual value). Data collected from IBGE, PNAD 2004~\cite{IBGE}} \label{fig:bra1}
\end{figure}
\begin{figure}
\includegraphics[width=0.7\textwidth]{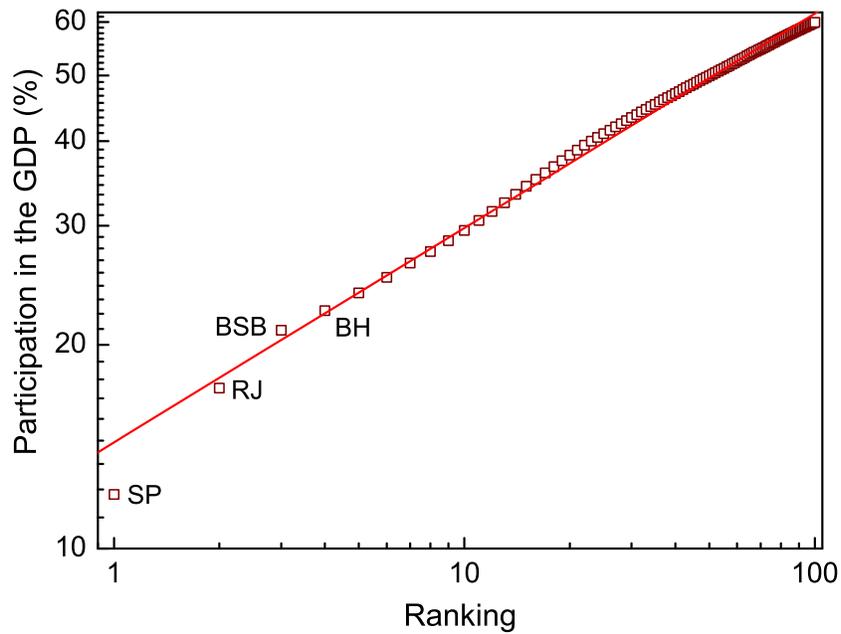}
\caption{Cumulative participation in the GDP of the $100$ richest counties in Brazil. The red line is an interpolation with a linear law, indicating, in the double logarithmic scale, that the distribution follows a power law. Data extracted from IBGE~\cite{IBGE2}} \label{fig:bra2}
\end{figure}

Power laws are not rare in nature, so it is not surprising that the wealth distribution follows a power law. The energy liberated in earthquakes, the size of avalanches, the size of cities, the frequency of all these phenomena are described by power laws~\cite{PerBak}. The quiz with the income and wealth distribution is not the power law, but how this distribution is generated through the dynamics of the agents interacting. On the other hand, a Pareto-Levy distribution is more unequal than a Gaussian: when the distribution follows a power law there are more affluent agents than in the case of a Gaussian distribution, but also more poor agents. And when the Pareto exponent increases the middle class tends to disappear.

In order to try to describe the processes that generate a given profile for the wealth distribution, in recent years diverse exchange models have been widely applied to describe wealth and/or income distributions in social systems. Different mathematical models of capital exchange among economic agents have been proposed trying to explain these empirical data (For a review see ref.~\cite{Caon2007}). Most of these models consider an ensemble of interacting economic agents that exchange a fixed or random amount of a quantity called ``wealth''.
This wealth represents the agents welfare. The exact choice of this quantity is not straightforward, but one can think that it stands for the exchange of a given amount of money against some service or commodity. Within these models the amount of exchanged wealth when two agents interact corresponds to some economic ``energy'' that may be randomly exchanged. If this exchanged amount corresponds to a random  fraction of one of the interacting agents wealth, the resulting wealth distribution is -- unsurprisingly -- a Gibbs
exponential distribution~\cite{dragu2000}.

Aiming at obtaining distributions with power law tails, several methods
have been proposed. Numerical procedures\cite{Caon2007,chatter1,chatter2,chakra,sinha2,IGVA2004}, as well as some analytical calculations~\cite{Bouchaud,cristian07}, indicate that one frequent result of that kind of models is condensation, i.e. concentration of all available
wealth in just one or a few agents. This result corresponds to a kind of equipartition of poverty: all agents (except for a set of zero measure) possess zero wealth while few ones concentrate all the resources. In any case, an almost ordered state is obtained, and this is a state of equilibrium, since agents with zero wealth cannot participate in further exchanges. The Gini coefficient~\cite{Gini} of this state is equal to 1, indicating perfect inequality~\cite{Caon2007}. Several methods have been proposed to avoid this situation, for instance, exchange rules where the poorer are favored~\cite{west,IGVA2004,Caon2007,sinha1,cristian07} but in all circumstances the final state is one with high inequality, i.e. very near condensation.

The exchange rules that produce condensation consider that, when two agents interact,  the exchanged amount $\Delta w$ is proportional to the wealth of one of the participants or to both~\cite{Bouchaud}. One particular example is $\Delta w = \mbox{min} \{ (1-\beta)w';(1-\beta)w''\}$, where  $w'$ and $w''$ are the wealth of the two interacting agents, and $\beta$ is the capital fraction that the agents risk during an exchange~\cite{Caon2007,IGVA2004,cristian07}. It is worth noting that even approaching a condensed state, in the intermediate stages the wealth distribution goes through a series of power law distributions where the Pareto exponent increases as a function of time~\cite{cristian07}. As real societies are not in a condensed state (since such a state would represent the ``thermal death'' of the economy), some kind of regulation must be present to guarantee that resources in the power law tail are re-injected back in the region of Gaussian distribution.

A few years ago we presented an alternative model for wealth distribution, the Conservative Exchange Market Model (CEMM), inspired by the ideas of John Rawls~\cite{Rawls} and Amartya Sen~\cite{Sen} and also on the Bak-Sneppen model for extinction of species~\cite{BakSnep}. The main point of the model is that some kind of action should be taken to change the state of the poorest agent in the society. The idea of a society that take measures in order to improve the situation of the most impoverished is compatible with the propositions of John Rawls, in his book ``A Theory of Justice''~\cite{Rawls}, directed towards an inventive way of securing equality of opportunity as one of the basic principles of justice. He asserts that {\it no redistribution of resources within a state can occur unless it benefits the least well-off: and this should be the only way to prevent the stronger (or richer) from overpowering the weaker (or poorer)}. The practical way to carry out this proposition in a simulation was adapted from the Bak-Sneppen model of Self-Organized Criticality applied to the extinction of biological species~\cite{BakSnep}. In this model the less fitted species disappears and is replaced by new one with different fitness, and the appearance of this new species affects the environment changing the fitness of the neighboring species. In 2003 we developed a similar model where the fitness is substituted by the assets of a particular agent, and the model is now conservative, the difference between the new and the old wealth is taken from (if positive) or given to (if negative) the assets of the neighbors of the poorest agent~\cite{PIAV2003,SI2004}. The distribution obtained follows an exponential law as a function of the square of wealth and a poverty line with finite wealth is also obtained, i.e. the poorer agents do not have zero wealth (as it happens in the most exchange models). Also, the Gini coefficient obtained is relatively low and compares well with the values of the Gini coefficient of some European countries as Denmark or Sweden~\cite{SI2004}. This suggest a path to decrease inequality in real societies~\cite{SI2004}.

Here we revisit this model including regulatory mechanisms in order to further diminish inequality. We will verify that proportional taxes have a important effect on redistribution, while ``uniform'' taxes, like consumption taxes, exhibit a lesser effect. But in both situations the impact on the poverty line and the Gini coefficient is really impressive.

In the next section we present a very short review of the original model: the Conservative Exchange Market Model (CEMM) and its main conclusions. Then, we introduce regulatory tools in the model in order to compare with the previous results: Two kind of regulatory mechanisms will be discussed and the obtained results will be compared. Finally, in the last section we present our conclusions.

\section{The Conservative Exchange Market Model - CEMM}
\label{CEMM}

The Conservative Exchange Market Model (CEMM)~\cite{PIAV2003} is a simple macroeconomic
model that consists of a one-dimensional lattice with N sites and periodic boundary conditions. That means that each site represents an economic agent (individuals, industries or
countries) linked to two neighbors. Periodic boundary conditions denote that the lattice closes on itself like a ring, being the last site in the chain neighbor of the first one. To each agent it is assigned some wealth-parameter that represents its welfare,
like the GDP for countries or accumulated wealth for individuals. One chooses an arbitrary initial configuration where the wealth is a number between 0 and 100 distributed randomly and uniformly among agents.
The dynamics of the system is supported on the idea that some measure should be taken
to modify the situation of the poorest agent. In this context, this process is simulated by
an extreme dynamics~\cite{BakSnep}: at each time step, the poorest agent, i.e., the one with the
minimum wealth, will perform (or be the subject of) some action trying to improve its economic state. Since the outcome of any such measure is uncertain, the minimum suffers a random change in its wealth, $\Delta w$~\cite{PIAV2003,SI2004}. In the first version of the model it is assume that whatever wealth is gained (or lost) by the poorest agent it will be at the expenses of its neighbors and that $\Delta w$ will be {\it equally} deducted from (or credited to) its two nearest neighbors on the lattice, making the total wealth constant. Numerical simulations on this model showed that, after a relatively long transient, the system arrives at a self-organized critical  state with a stationary wealth distribution (Fig. 1 of Ref.~\onlinecite{PIAV2003}) in which almost all agents are above a certain threshold or poverty line.

Another possibility is to subtract $\Delta w$ from two agents picked at random. This situation has also been considered~\cite{IGPVA2003,SI2004}, it is the {\it annealed} or mean field version of the model, and corresponds to a situation in which the agents with which the exchange takes place are chosen at random and not based on geographical proximity.

In the nearest-neighbor version of the model, one founds a minimum wealth or poverty
line that is $\eta_T \approx 40 \%$ of the maximum initial wealth and above this threshold the distribution of the wealth of the agents is an exponential $P(w) \approx exp(-w^2/2 \sigma^2)$, with $\sigma=22.8$~\cite{IGPVA2003}. The obtained Gini coefficient, $G$~\cite{Gini}, is very low, of the order of $G=0.1$. In the mean-field case the model exhibits a lower threshold, $\eta_T \approx 20 \%$, and, beyond it, also an exponential distribution with a higher value of $\sigma \approx 56.7$~\cite{IGPVA2003} and of the Gini coefficient $G \approx 0.25$.

In the present article we would like to discuss a similar model still using extremal dynamics, but making some changes in the way the wealth is redistributed. In the original CEMM model~\cite{PIAV2003,SI2004} the quantity of money of the agent with the minimum wealth is changed to an arbitrary value between the limits $\{0,100\}$ and the difference between the new and the old wealth, $\Delta w$, is taken from the neighbors in the local version of the model or from agents picked at random in the mean-field version.

Now, we will include regulations on the form of taxes, and we will consider two scenarios: 1) Uniform taxes and 2) Proportional taxes.

\section{CEMM with uniform taxes}
\begin{figure}
\includegraphics[scale= 1.0,angle=0]{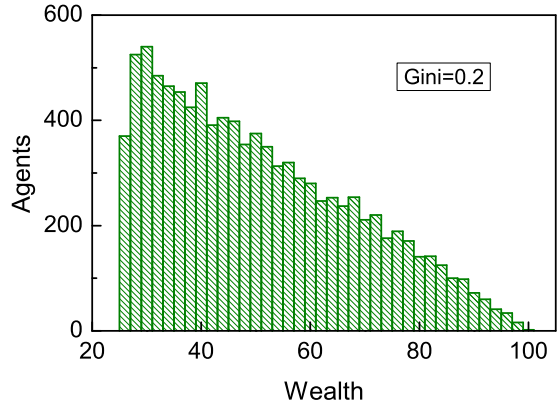}
\caption{Plot of  the wealth distribution for $N=10000$ agents considering that taxes are uniformly distributed among all agents. The poverty line is $\eta_T=25$ } \label{fig:tax1}
\end{figure}
 Let us first consider the situation where the amount attributed to the poorest agent is equally collected from the full population. In this way each of the agents (including the agent of minimum wealth) will contribute to improve the situation of the less favored ones. This kind of redistribution simulates a tax that is the same for every agent. Taxes on consumption, like TVA in Europe, ICM in Brazil, IVA in Argentine, etc., are of this type, every agent contributes the same amount independent of his available resources. The obtained wealth distribution in this case of uniform and global taxes, is of the same type as in the original model~\cite{PIAV2003} but now the poverty line is lower than in the local case ($\approx 25\%$ of the maximum wealth) and higher that in the mean-field case. Also, the Gini coefficient is in between $G \approx 0.2 $. Finally the wealth distribution is almost linear and it is represented in Fig. 3, where we show the results for a number of agents $N=10^4$ and $10^7$ time-steps (we have verified that in this limit the system has attained the state of self-organize criticality  and there are no more changes neither in the poverty line nor in the Gini coefficient). So, a system with uniform taxes exhibit less inequality than the previous mean-field model, but higher than the local model.

\section{CEMM with Proportional Taxes}

Let us now assume a more equitable way to apply taxes: proportional taxes, equivalent to the taxes on income or on fortune. Here we will assume a linear proportionality, but it is very easy to modify the model to consider progressive or regressive taxation. As this case has not been previously studied we will consider both the local and global scenarios.

\subsection{Local taxes}

\begin{figure}
\includegraphics[scale= 1.0,angle=0]{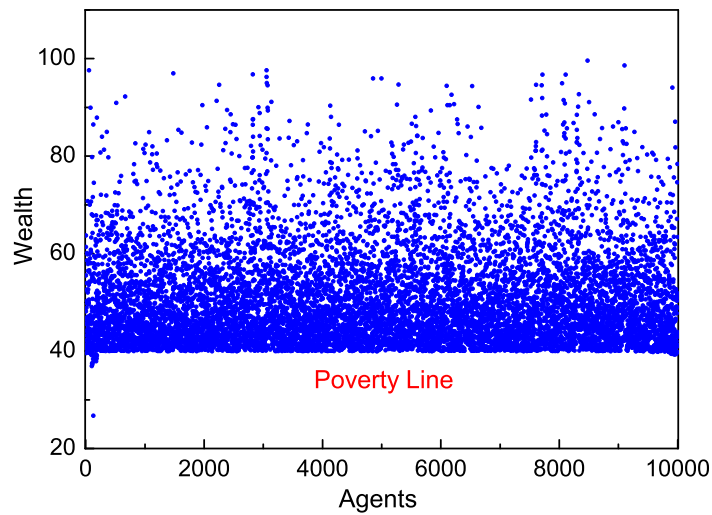}
\caption{Plot of  the wealth of each agent in a system with $10000$ agents with local proportional taxes. ``Islands'' of poverty and affluence are observed, but the poverty line is high, of the order of $40$} \label{fig:tax2}
\end{figure}
\begin{figure}
\includegraphics[scale= 1.0,angle=0]{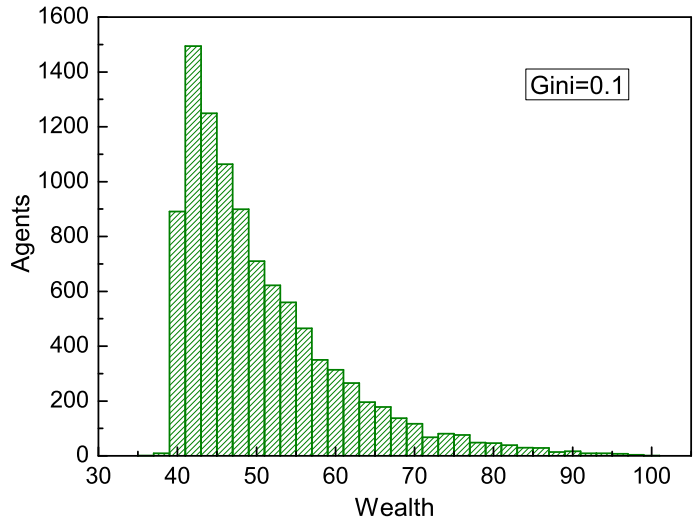}
\caption{Histogram of the wealth distribution with local proportional taxes. There is a strong middle class and the distribution is exponential, as if refs.~\onlinecite{PIAV2003,IGVA2004}} \label{fig:tax3}
\end{figure}
The dynamics here is very similar to the original CEMM. In the original model if the agent with minimum wealth changes his fortune by a quantity $\Delta w$, this capital (positive or negative) is equally subtracted from his neighbors. Each one of them have a deduction  $\Delta w/p$ where $p$ is the number of neighbors considered\cite{IGVA2004}. We will consider now that the deduction is not equally performed but proportional to the wealth of the neighbor.
We define a factor $\lambda=\Delta w/K_{nn}$ where $K_{nn}$ is the total capital of the $p$ neighbors. Each neighbor will suffer a subtraction (if $\Delta w > 0$) or addition (if $\Delta w < 0$) on his wealth equal to $\lambda w_i$. That means that if the i-agent is one of the neighbors his wealth at time $(t+1)$ will be
\begin{equation}
w_i(t+1)=(1-\lambda) w_i(t),
\end{equation}\label{eqn:lambda}

\noindent i.e., his wealth will be reduced if $\Delta w > 0$ or will increase if $\Delta w <0$. It is straightforward to verify that with this condition the total wealth is conserved.

In Fig. 4 we have represented the wealth of the agents in a scale from 0 to 100. We have again considered $10^4$ agents and $p=4$ neighbors. Because the transfer of wealth is local, some ``geographical'' segregation appears in the wealth distribution. This is clear in Fig. 4 where it is possible to see that the poverty line is relatively higher than in the previous situation, of the order of $40$, but also that there are geographical differences: on the left one observes a ``burst'' of poor agents below the poverty line, but it is also possible to see that there are wealthy and middle class regions in the plot, indicating that the dynamics induces regional differences. Certainly this kind of behavior should be visible in a more emphatic way if one consider a complex lattice instead of the very simple one-dimensional model that we describe here. In any case, the Gini coefficient is very low, of the order of $G=0.1$, as for the local model previously studied~\cite{IGPVA2003}. Therefore, even if the results are encouraging they are not very different from the original CEMM where the difference received by the poorest agent were equally shared (and not proportional) by the neighbors. Also, the wealth distribution is exponential, as in ref.~\cite{IGVA2004} and it is represent5ed in Fig. 5: there is a peak in the number of agents just above the poverty line, and then the number of agents decreases exponentially as the wealth increases.

It is when one examines the application of global taxes that differences appear, as well in the poverty line and the Gini coefficient than in the type of distribution.

\subsection{Global taxes}

Let us now consider that the amount $\Delta w$ received by (or deduced from) the poorest agent is subtracted from (or added to) all the agents (including the recipient) but in proportion to their wealth. Equation (1) is still valid but now $\lambda=\Delta w/K_{tot}$, being $K_{tot}$ the wealth of the full society.

Globalization has its shortcomings when compared to local models but still the results are impressive. The poverty line is a little lower than in the local case, $32$ compared to $40$ and the Gini coefficient higher, of the order of $G=0.16$. We have represented on Fig. 6 the wealth of each agent and we can verify that the distribution is more homogeneous from a geographical point of view. However, if one examines the wealth distribution, plotted on Fig. 7, the figure suggest a power law distribution, even if we do not have enough statistics, particularly for the tail of affluent agents, to determine the value of the exponent. In any case the distribution is not exponential and this is an expected when a multiplicative redistribution is considered.
Finally if we compare the distributions for local taxes (Fig. 5) and global taxes (Fig. 7) it evident that the number of poor agents is lower and the number of middle class agents is higher in the situation with global taxes, while the poverty line is higher in the case of local taxes.

Then when proportional taxes are applied to the full population, each agent contributes with a very small amount, but we obtain a society where there are no zero-wealth agents, neither geographical disparities, as it can be observed in Fig. 6. Also, the wealth distribution is a power law, then the number of affluent agents is bigger than in the case of an exponential distribution, nevertheless the Gini coefficient is still very low, as it is shown on Fig. 7.

\begin{figure}
\includegraphics[scale= 1.0,angle=0]{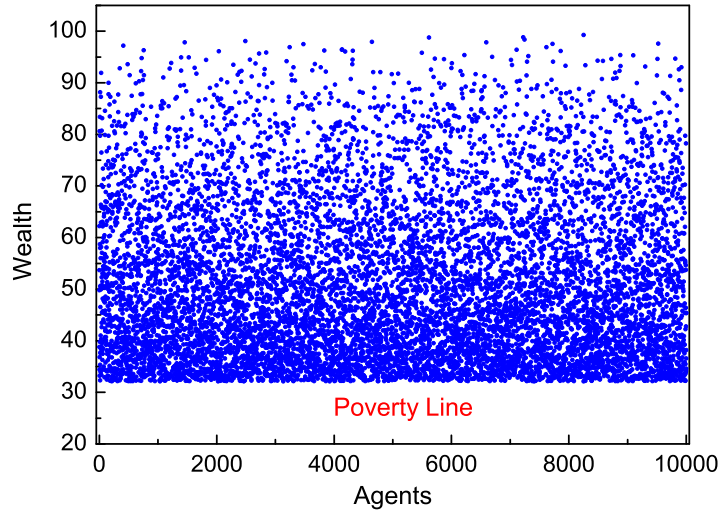}
\caption{Plot of  the wealth of each agent in a system with $10000$ agents, for the case of global proportional taxes. The distribution is uniform, without islands, and the poverty line is of the order of $32$} \label{fig:tax2}
\end{figure}
\begin{figure}
\includegraphics[scale= 1.0,angle=0]{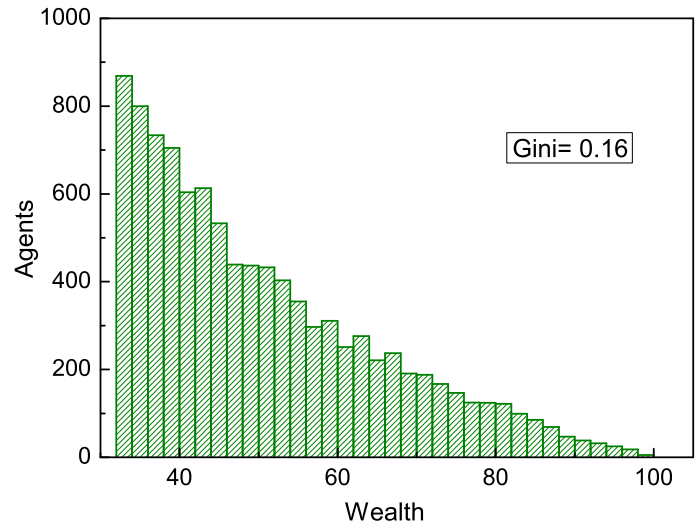}
\caption{Histogram of the wealth distribution in the case of global proportional taxes. There is still a strong middle class and the distribution looks like a power law} \label{fig:tax3}
\end{figure}

\section{Discussion and Conclusion}

Inequalities are related to social and economic phenomena and its origin is a subject
of much debate. The fraction of rich and poor people in a country depends on the
inequalities of the income distribution but also on the remuneration of labor, profits, taxes, etc.
The model here presented is certainly a very simplified one; for example, taxes do not play just a redistributive objective, but also are used by governments to assure the infrastructure needed to economic development. Nonetheless, in spite on its simplicity the present model describes a very strong redistributive mechanism, that coincides with some public policies, like the ``bolsa-familia'' (family fellowship, allocate to very poor family groups) in Brazil, or the small loans to jobless people in order to develop their own business. Even in the global case, and with taxes proportional to the possessions of each agent, the contribution of each one of them is very small but results in a low Gini coefficient, even lower than the one observed in the most egalitarian countries, like Denmark or Japan (where the Gini coefficient is of the order of $G=0.25$). Finally, the poverty line is much higher that the one obtained with purely exchange models.

We remark also that the local version of the model generates a wealth distribution with a very low Gini index, so very close to an full equitable society, but with geographical differences. In the global case, the Gini coefficient is a little higher , but from a geographical point of view the distribution is more uniform, so maybe this is a point in favor of globalization. And coming back to the inspiration of this work, it is interesting that the minimum dynamics favors wealth redistribution when acting on the poorest agents, in the same sense defined by Rawls~\cite{Rawls}. It seems that this kind of dynamics, when applied to markets, is a road to ensure to the poorest agent a chance to improve his situation.

We conclude that the model, in spite of its simplicity, is able to reproduce some properties of modern economies, particularly the wealth distribution of welfare societies, and it indicates a way to improve the situation of extreme inequality in some countries with very high Gini indexes.

\section*{Acknowledgements}
We acknowledge fruitful discussions with G. Abramson, S. Gon\c{c}alves,
F. Laguna, S. Souza, R. Don\'angelo and J.L. Vega. We acknowledge financial support from Brazilian agencies CNPq and CAPES.


\end{document}